\documentclass[aps,reprint,nofootinbib,nobibnotes,notitlepage,superscriptaddress,twocolumn,prl,
 amsmath,amssymb
]{revtex4-2}

\usepackage[caption=false]{subfig}
\usepackage{braket}
\usepackage{float}
\usepackage{lipsum}
\usepackage{graphicx}
\usepackage{dcolumn}
\usepackage{bm}
\usepackage{natbib}
\usepackage{hyperref}
\hypersetup{
	colorlinks = true,
    linkcolor = Red,
    urlcolor  = Red,
    citecolor = Red
}

\usepackage{microtype}

\usepackage{verbatim}
\usepackage[amssymb]{SIunits}
\usepackage{tabularx}
\usepackage[dvipsnames]{xcolor}
\usepackage{wasysym}

\usepackage{tikz}

\def\be{\begin{equation}}
\def\ee{\end{equation}}

\usepackage{color}

\definecolor{darkgreen}{RGB}{0,120,0}

\newcommand{\Mpch}{h^{-1}\mathrm{Mpc}}

\newcommand{\av}[1]{\left\langle{#1}\right\rangle}

\newcommand{\vx}{\vec x}

\newcommand{\C}{\mathsf{C}}

\newcommand{\hr}{\hat{\vec r}}

\renewcommand{\vr}{\vec r}

\renewcommand{\P}{\mathcal{P}}

\def\beq{\begin{eqnarray}}
\def\eeq{\end{eqnarray}}
\let\vec\mathbf
\usepackage{xspace}
\newcommand*{\patchy}{\selectfont\textsc{MultiDark-Patchy}\xspace}
\newcommand*{\glam}{\selectfont\textsc{GLAM-Uchuu}\xspace}

\usepackage{empheq}

\def\mnras{\rm{MNRAS}}
\def\aj{\rm{AJ}}

\def\apj{\rm{ApJ}}                 
\def\jcap{\rm{JCAP}}

\begin{document}

\title{Could Sample Variance be Responsible for the Parity-Violating Signal Seen in the BOSS Galaxy Survey?}

\author{Oliver~H.\,E.~Philcox}
\email{ohep2@cantab.ac.uk}
\affiliation{Department of Physics,
Columbia University, New York, NY 10027, USA}
\affiliation{Simons Society of Fellows, Simons Foundation, New York, NY 10010, USA}

\author{Julia Ereza}
\email{jferrer@iaa.es}
\affiliation{Instituto de Astrofísica de Andalucía (CSIC), Glorieta de la Astronomía, E-18080 Granada, Spain}

\begin{abstract} 
    \noindent Recent works have uncovered an excess signal in the parity-odd four-point correlation function measured from the BOSS spectroscopic galaxy survey. If physical in origin, this could indicate evidence for new parity-breaking processes in the scalar sector, most likely from inflation. At heart, these studies compare the observed four-point correlator to the distribution obtained from parity-conserving mock galaxy surveys; if the simulations underestimate the covariance of the data, noise fluctuations may be misinterpreted as a signal. To test this, we reanalyse the BOSS CMASS \& LOWZ parity-odd dataset with the noise distribution modeled using the newly developed \textsc{GLAM-Uchuu} suite of mocks. These comprise full $N$-body simulations that follow the evolution of $2000^3$ dark matter particles in a $\Lambda$CDM universe, and represent a significant upgrade compared to the formerly used \textsc{MultiDark-Patchy} mocks, which were based on an alternative (non $N$-body) gravity solver. We find no significant evidence for parity-violation in the BOSS dataset (with a baseline detection significance of $1.4\sigma$), suggesting that the former signal ($>$$3.5\sigma$ with our data cuts) could be caused by an underestimation of the covariance in \textsc{MultiDark-Patchy}. The significant differences between results obtained with the two sets of BOSS-calibrated galaxy catalogs showcases the heightened sensitivity of beyond-two-point analyses to the treatment of non-linear effects and indicates that previous constraints may suffer from large systematic uncertainties.
\end{abstract}

\maketitle

\noindent Is the large-scale Universe invariant under a parity transformation? According to the standard model of cosmology, the answer is yes, but a slew of recent observations have challenged this notion. In particular, several works have obtained evidence for \textit{birefringence} in the \textit{Planck} cosmic microwave background (CMB) polarization \citep[e.g.,][]{Eskilt:2022wav,Eskilt:2022cff,Diego-Palazuelos:2022dsq,Komatsu:2022nvu,Minami:2020odp,Diego-Palazuelos:2022cnh,Clark:2021kze}, which could indicate novel tensorial physics such as an axion-photon coupling \citep{Turner:1987bw}. Moreover, analyses of the spectroscopic galaxy four-point correlation function (hereafter 4PCF) have found an excess scalar parity-violating signal \citep{Philcox:2022hkh,Hou:2022wfj} (using methods first discussed in \citep{Cahn:2021ltp}). Whilst it seems unlikely that these effects are related, each could be an intriguing hint of physics beyond the standard paradigm; however, one must keep in mind the possibility that the signatures could be non-physical in origin. In this \textit{Letter}, we focus on the galaxy parity-violation measurements, with the hope of disentangling novel phenomena from analysis systematics.

To constrain parity-violation with the distribution of galaxies (which transforms as a scalar), one requires statistics based on the 4PCF or beyond. This occurs since tetrahedra (which describe the vertices of a quadruplet of galaxies) are the first chiral polyhedron, and thus the simplest shape for which one can distinguish reflection and rotation (as discussed in \citep{Cahn:2021ltp}, see also \citep{Lue:1998mq,Jeong:2012df,Shiraishi:2016mok,Gluscevic:2010vv,Liu:2019fag}). Roughly speaking, parity-violation measurements proceed by iterating over all possible tetrahedra of galaxies in a survey and counting the number that are left- and right-handed as a function of shape (which is a 4PCF in disguise).\footnote{This can be done uniquely; one considers the sign of the triple product $\vr_1\cdot\vr_2\times \vr_3$, where $\vr_i$ is the position vector from the zeroth vertex to the $i$-th and we fix $r_1< r_2< r_3$.} Efficient methods for computing this exist \citep[e.g.,][]{npcf_algo,Philcox:2021eeh,gen_npcf} (using mathematical tools developed in \citep{Cahn:2020axu}), which have facilitated parity-violation measurements from the SDSS-III BOSS galaxy survey containing $\mathcal{O}(10^6)$ galaxies \citep{2013AJ....145...10D,2017MNRAS.470.2617A}.

In essence, the analyses of \citep{Philcox:2022hkh} and \citep{Hou:2022wfj} proceed by computing the parity-odd 4PCF of BOSS data and interpreting this result in terms of the expected distribution arising from cosmic variance and shot-noise. Both works report significant deviations from the null distribution; \citep{Philcox:2022hkh} find a $2.9\sigma$ excess in the CMASS sample, whilst \citep{Hou:2022wfj} obtained a $7.1\sigma$ ($3.1\sigma$) deviation in the CMASS (LOWZ) sample. These analyses differ in their approach and assumptions; \citep{Philcox:2022hkh} primarily makes use of non-parametric rank tests, whilst \citep{Hou:2022wfj} utilize $\chi^2$ tests, facilitated either by rescaling a theoretical covariance or projected the data into a low-dimensional subspace; furthermore, they find larger signals when adopting finer radial binning. 

Analysis differences notwithstanding, both \citep{Philcox:2022hkh} and \citep{Hou:2022wfj} find an intriguing excess in the 4PCF; if physical, this would suggest parity-violating processes at work in the Universe, which could be of relevance to both the observational and theoretical communities. For the former, we note that searches for parity-violation have also been performed using the CMB four-point function \citep{PhilcoxCMB,Philcox:2023ypl} (and the tensorial three-point function \citep{Philcox:2023xxk}), and there remains the intriguing possibility that such effects could be searched for in galaxy shape and spin statistics \citep[e.g.,][]{Philcox:2023uor,Coulton:2023oug,Jia:2022sys,Yu:2019bsd,Motloch:2021mfz,Kogai:2020vzz}. To date, none of these studies have reported a significant detection. On the theory side, a wide variety of works have considered parity-violating models both in the scalar and tensor sector \citep{Alexander:2009tp,Alexander:2011hz,Alexander:2016hxk,CyrilCS,Bartolo:2017szm,Bartolo:2015dga,Sorbo:2011rz,Shiraishi:2014ila,Shiraishi:2014roa,Shiraishi:2016mok,Alexander:2006mt,Gong:2023kpe,Soda:2011am,Bordin:2020eui,Cabass:2022oap,Cabass:2022rhr,Shiraishi:2012sn,Shiraishi:2011st,Bartolo:2014hwa,Shiraishi:2013kxa,Liu:2019fag,Masui:2017fzw}. Applied to BOSS data, \citep{Cabass:2022oap} drew the general conclusions: (a) late-time solutions (\textit{i.e.}\ non-inflationary) are heavily suppressed on large-scales, (b) none of the theoretical models yet analysed are consistent with the observational data. 

Perhaps the least exciting explanation for the BOSS parity-odd-excess is a mischaracterization of the 4PCF noise distribution. Both previous analyses essentially analyze the following quantity:
\beq\label{eq: chi^2}
    \tilde{\chi}^2 = \zeta_{\rm odd}^T\C^{-1}\zeta_{\rm odd},
\eeq
where $\zeta_{\rm odd}$ is the (possibly compressed) 4PCF and $\C$ is some covariance matrix, calibrated to a suite of realistic galaxy catalogs, namely the \patchy suite \citep{2016MNRAS.456.4156K}. In the Gaussian limit, and assuming $\C$ is the true covariance of $\zeta$, $\tilde\chi^2$ follows a $\chi^2$ distribution (with the null theory $\zeta_{\rm odd} = 0$), thus the value of $\tilde\chi^2$ obtained from BOSS can be interpreted as a parity-violation detection probability. Of course, the actual analyses are somewhat more nuanced, and take into account effects such as the finite number of mocks used to estimate $\C$ and non-Gaussianities in the distribution of $\tilde\chi^2$ (via a simple invocation of simulation-based-inference) \citep{Philcox:2022hkh,Hou:2022wfj}. At heart, all analyses make a key assumption: \textit{that the noise distribution of $\zeta_{\rm odd}$ from BOSS matches the noise distribution of $\zeta_{\rm odd}$ from the \patchy mock catalogs}. 

In this \textit{Letter}, we test the above assumption. This is facilitated by a new suite of BOSS mocks: the \glam galaxy catalogs \citep{Ereza:2023zmz}. These are $N$-body cosmological simulations that follow the evolution of $2000^3$ dark matter particles, each having a particle mass of $1.06\times10^{10}\,h^{-1}\mathrm{M}_\odot$. The cosmological parameters adopted are $\Omega_\mathrm{m,0}=0.309$, $\Omega_\mathrm{b,0}=0.0486$, $\Omega_\mathrm{\Lambda,0}=0.691$, $h=0.677$, $n_\mathrm{s}=0.9667$, and $\sigma_8=0.816$, representing the best fitting $\Lambda$CDM parameters corresponding to the \textit{Planck} 2015 cosmology \citep{Planck:2015fie}.
In contrast, the previously used \patchy catalogs, rely on an approximate model for gravitational evolution, rather than employing $N$-body codes. The \textsc{Patchy} code generates fields for dark matter density and peculiar velocity on a mesh, utilizing Gaussian fluctuations and implementing the augmented Lagrangian perturbation theory scheme \citep{2014MNRAS.439L..21K}.

Whilst alternative methods to $N$-body simulations are more computationally efficient, they necessarily sacrifice some accuracy, implying that the \patchy simulations fail to generate a precise matter density field compared to a full $N$-body simulation such as \glam, which fully encodes nonlinear gravitational evolution (up to resolution and baryonic effects). The feasibility of the resulting \patchy galaxy catalogs in a real universe is uncertain, which introduces ambiguity in their covariance error estimates for the BOSS clustering statistics. In addition, the generation of these catalogs involves specifying a significant number of parameters (five), leading to a remarkable degree of parameter degeneracy. In contrast, \glam has only one free parameter. A further important difference is that, unlike \patchy, \glam includes the redshift-dependent evolution of galaxy clustering. Whilst the \patchy catalogs were generated using a single $ 2.5\,h^{-1}\mathrm{Gpc}$ box, the \glam catalogs were generated by joining boxes of different redshift into spherical shells, reproducing the clustering evolution with redshift (see \citep{Ereza:2023zmz} for a detailed description of this method). Moreover, \citep{Yu_2023} recently reported some evidence of model specification errors in \patchy, and \citep{Ereza:2023zmz} found differences in the two-point covariances.

In the below, we perform a similar analysis to \citep{Philcox:2022hkh}, but replacing \patchy with the \glam galaxy catalogs. Assuming that they better characterize the null $\zeta_{\rm odd}$ distribution (which is a fair assumption, \textit{viz} the above discussion), this will allow us to test whether the formerly-reported BOSS 4PCF excess could be explained by a simple underestimate of the data's sample variance.

\section{Methods}
\noindent Our observational data-set is the twelfth data release (DR12) of the Baryon Oscillation Spectroscopic Survey (BOSS), part of SDSS-III \citep{2013AJ....145...10D,2017MNRAS.470.2617A,2011AJ....142...72E}. As in \citep{Hou:2022wfj} (but extending beyond \citep{Philcox:2022hkh}) we use both the low-redshift LOWZ sample ($0.2<z<0.4$) and the high-redshift CMASS sample ($0.43<z<0.7$), which represent two physically distinct galaxy populations, and split the data into the North and South galactic regions (denoted `N' and `S' hereafter). Adopting the standard BOSS systematic weights \citep{2018MNRAS.477.1153V,BOSS:2016apd} (which we caution have been validated only for the two-point functions), the effective number of galaxies in our sample is $606\,000$, $225\,000$, $196\,000$ and $91\,103$ for CMASS-N, CMASS-S, LOWZ-N and LOWZ-S respectively.

We use two sets of mock catalogs in this study: \patchy and \glam. Both model the full BOSS footprint (split across all four chunks) with similar redshift-distributions and observational masks. Here, we use $N_{\rm sim}=2042$ \patchy and $N_{\rm sim}=627$ \glam mocks.\footnote{$6$ \patchy and $3$ \glam mocks were corrupted in data transfer and are thus dropped from the analysis.} Both sets of catalogs include the observational veto mask and estimated fiber collision weights, emulating the observational sample. We additionally utilize random catalogs to characterize the unperturbed galaxy distribution (from the mask and radial distribution). These have been separately generated for each dataset (BOSS, \glam and \patchy) and have size $40\times$, $30\times$, and $20\times$ the data respectively.

\begin{figure}
    \centering
    \includegraphics[width=\linewidth]{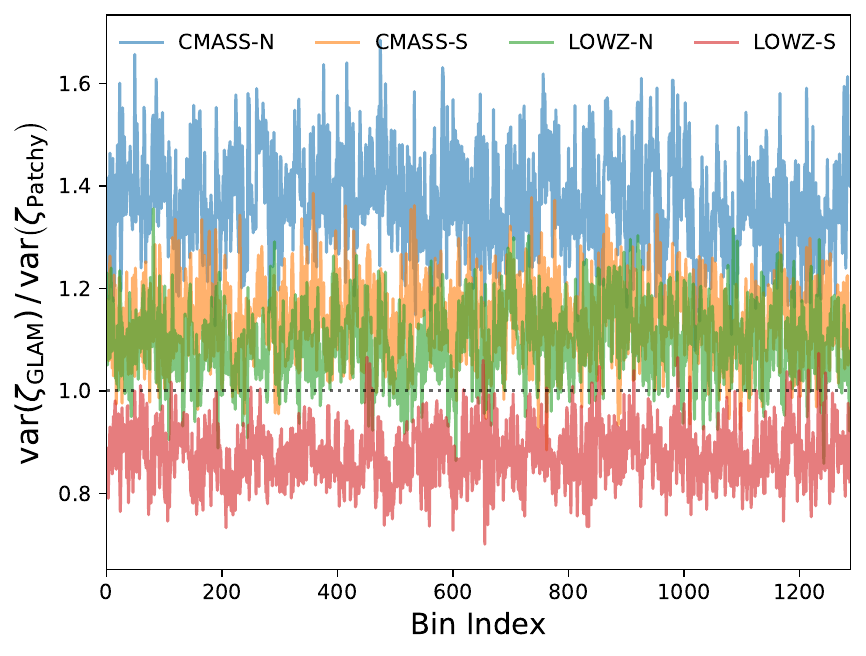}
    \caption{Comparison of the parity-odd 4PCF variance estimated with two suites of galaxy mock catalogs: \glam ($N$-body) and \patchy (mesh, with augmented Lagrangian perturbation theory). We show results for the four BOSS samples, as indicated in the caption, and collapse all $1288$ 4PCF bins into one dimension. In all samples except LOWZ-S, the \glam variance exceeds that of \patchy. This gives rise to the differences seen in Figs.\,\ref{fig: rank-test}\,\&\,\ref{fig: rank-test-individual}.}
    \label{fig: cov-ratio}
\end{figure}

Given the datasets, the 4PCFs are obtained using the \textsc{encore} code \citep{npcf_algo,Philcox:2021hbm}. This computes the following quantity (with a complexity quadratic in the number of galaxies):
\beq\label{eq: 4pcf-def}
    \zeta_{\ell_1\ell_2\ell_3}(r_1,r_2,r_3) &=& \int d\vx\, d\hr_1d\hr_2d\hr_3\P_{\ell_1\ell_2\ell_3}^*(\hr_1,\hr_2,\hr_3)\\\nonumber
    &&\,\times\,\delta(\vx)\delta(\vx+\vr_1)\delta(\vx+\vr_2)\delta(\vx+\vr_3),
\eeq
where $\delta$ is the galaxy overdensity,\footnote{In practice, we use the data and random catalogs to create a Landy-Szalay-type estimator \citep{1993ApJ...412...64L} to remove the window function, as described in \citep{npcf_algo,Philcox:2021hbm}.} and $\P_{\ell_1\ell_2\ell_3}$ is an angular basis function relative to one vertex of the galaxy tetrahedron (as defined in \citep{Cahn:2020axu}; see also the BiPoSH basis \citep{1988qtam.book.....V}). The $r_i$ parameters give the (discretized) tetrahedron side lengths, whilst $\ell_i$ index the Fourier complement of the internal angle, with odd $\ell_1+\ell_2+\ell_3$ encoding parity-violation. Our binning follows \citep{Philcox:2022hkh}; we use all $\ell$ up to $\ell=5$, but drop $\ell=5$ due to residual leakage from the window function, and use $10$ radial bins with $r_i\in [20,160)\Mpch$, additionally dropping any configurations with internal radii (\textit{i.e.}\ the three tetrahedron sides not specified by $r_{1,2,3}$) below $14\Mpch$.

\begin{figure}
    \centering
    \includegraphics[width=0.9\linewidth]{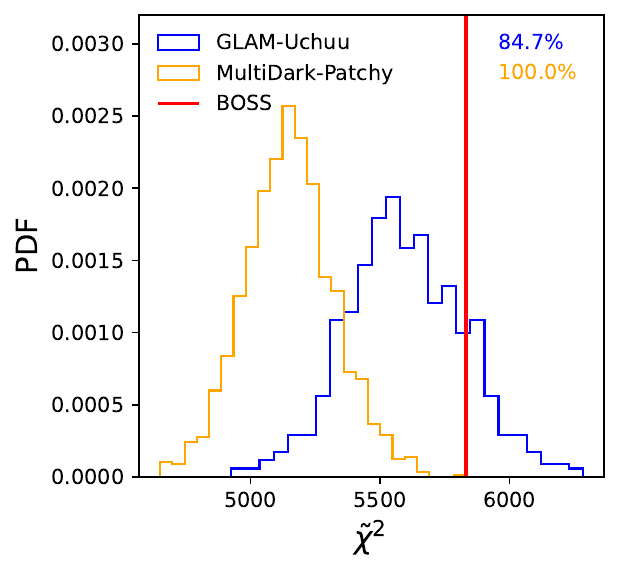}
    \caption{Distribution of the parity-odd statistic $\tilde{\chi}^2$ \eqref{eq: chi^2} from 2042 \patchy catalogs (used in previous analyses) and 627 \glam catalogs (new to this work), with the BOSS result shown as a vertical red line. We show results from the combined CMASS+LOWZ observational dataset, with those from the four subsamples displayed in Fig.\,\ref{fig: rank-test-individual}. The rank-test probabilities (\textit{i.e.}\ detection significances) are shown in the upper right corner. Using the \glam catalogs, we find no evidence for parity-violation in BOSS, but a strong preference when the 4PCF noise distribution is modeled with the \patchy suite. Further, we find $\geq$$3\sigma$ evidence for parity-violation in $41\%$ of the (parity-conserving) \glam mocks, when analyzed with \patchy. Numerical constraints are given in Tab.\,\ref{tab: rank-probs}.}
    \label{fig: rank-test}
\end{figure}

\begin{figure*}
    \centering
    \includegraphics[width=0.95\textwidth]{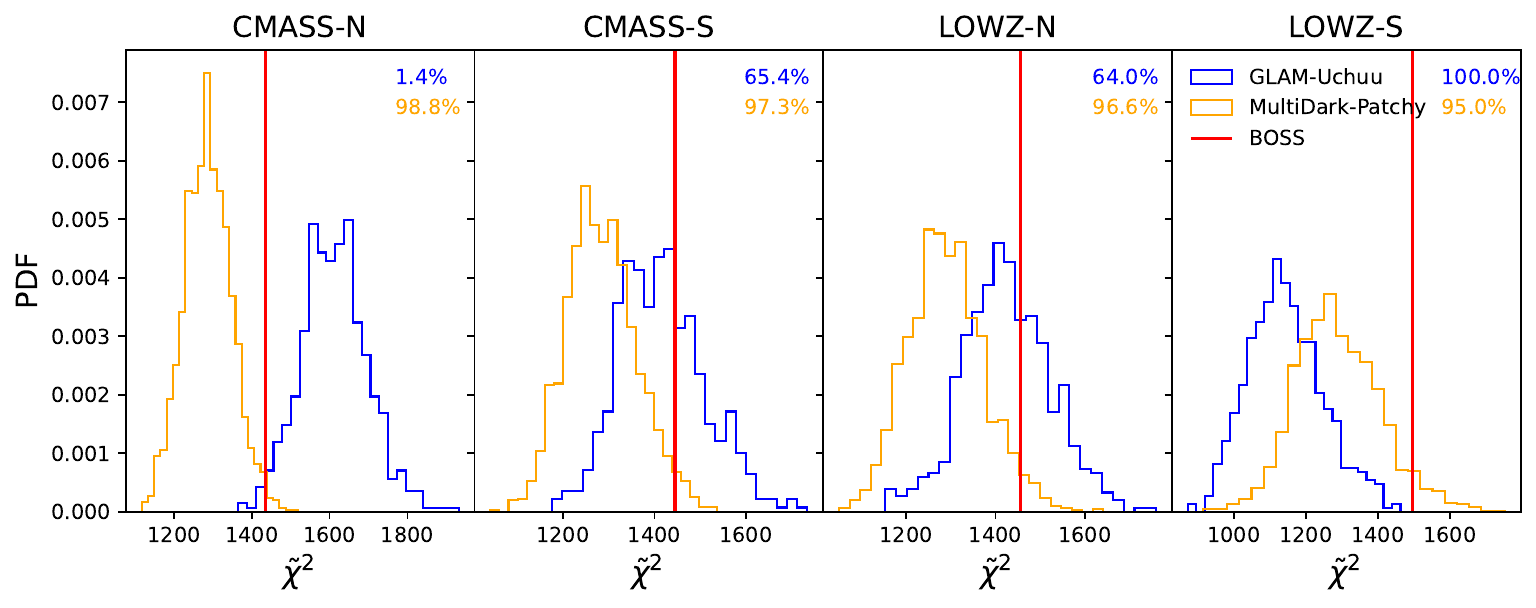}
    \caption{As Fig.\,\ref{fig: rank-test}, but assessing the probability of parity-violation in each of the four BOSS data chunks (indicated by the title). The PDFs of \glam and \patchy are significantly shifted in all except LOWZ-N, as expected from Fig.\,\ref{fig: cov-ratio}. We find no detection of parity-violation in three of the four samples (see Tab.\,\ref{tab: rank-probs}), though the LOWZ-S result is anomalous, suggesting possible underestimated covariances or observational systematics.}
    \label{fig: rank-test-individual}
\end{figure*}

Computation of the 4PCFs was performed on a high-performance computing cluster. Thanks to \textsc{encore}'s efficient \textsc{C++} and \textsc{avx} implementation, $\approx 90$ minutes was needed to process each mock (including all four data chunks) on a 40-CPU machine, thus the full analysis required $\approx 16\,000$ CPU-hours.\footnote{Though the \textsc{encore} GPU implementation described in \citep{npcf_algo} is considerably faster, we do not use it here due to the far larger number of CPUs available on a typical cluster.} The full data-vector, $\zeta_{\rm odd} = \{\zeta_{\ell_1\ell_2\ell_3}(r_1,r_2,r_3)\,\forall\,r_i,\ell_i\}$ contains $N_{\rm bins}=1288$ bins for each of the four (assumed independent) data chunks.

To analyze the data, we follow the methods of \citep{Philcox:2022hkh}, making use of a fiducial analytic covariance matrix $\C$ computed as described in \citep{Hou:2021ncj}. This is estimated using various simplifying assumptions (in the Gaussian limit, for example), thus we use it only as an approximate tool for dimensionality reduction (noting that former works \citep{Hou:2021ncj,Philcox:2021hbm,Philcox:2022hkh} have found that it well-represents the empirical correlation structure of the 4PCF, but not the variance). On each of the data and mocks, we estimate a \textit{pseudo}-$\chi^2$ statistic (hereafter $\tilde\chi^2$) via \eqref{eq: chi^2}. If $\C$ is close to the true covariance, this will provide an optimal noise weighting in the Gaussian limit. Given the set of $\tilde\chi^2$ values obtained from a suite of (parity-conserving) mock catalogs, we construct an empirical distribution $p_{\rm sim}(\tilde\chi^2)$ and assess the probability-to-exceed of BOSS, \textit{i.e.}\ $p_{\rm sim}(\tilde\chi^2_{\rm BOSS})$. This is practically implemented as a rank test, which is a simple form of simulation-based inference. Since we take the distribution of $\tilde\chi^2$ directly from the catalogs, we do not need to assume Gaussianity nor that $\C$ well-approximates the true covariance; however, we must assume that $p_{\rm sim}(\tilde\chi^2)$ is accurate, \textit{i.e.}\ that the simulated noise matches the observational noise. When analyzing multiple data chunks, we sum the $\tilde\chi^2$ values of each, weighted by the ratio of the fiducial covariance $\C$ to the empirical one, as estimated from 10\% of the \patchy catalogs using a Kullback-Leibler minimization (\textit{i.e.}\ inverse noise weighting). Once again, we note that this cannot induce bias.

We additionally perform a second test for parity-violation, used in both \citep{Philcox:2022hkh} and \citep{Hou:2022wfj} (following \citep{Philcox:2021hbm,1999ApJ...517..531S}). Given the high-dimensionality of $\zeta_{\rm odd}$, we cannot compute the empirical covariance $\C_{\rm sim} = \av{\zeta_{\rm odd}^{\,}\zeta_{\rm odd}^T}_{\rm sim}$ directly from the mock catalogs. Instead, we use the fiducial covariance $\C$ to define a minimum-variance eigenbasis containing $N_{\rm eig}=250$ elements, into which the data and mock 4PCFs are projected via some $N_{\rm eig}\times N_{\rm bins}$ operator $\Pi$.\footnote{The choice of $N_{\rm eig}$ is somewhat arbitrary, but \citep{Philcox:2022hkh} found $N_{\rm eig}=250$ to yield large detection significances with \patchy catalogs whilst limiting finite-mock effects.} This is then analyzed with the following statistic:
\beq\label{eq: T2}
    T^2 = (\Pi\zeta_{\rm odd})^T\C_{\rm sim}^{-1}(\Pi\zeta_{\rm odd}),
\eeq
which is closely related to \eqref{eq: chi^2} but uses the empirical covariance  $\C_{\rm sim} = \av{(\Pi\zeta_{\rm odd})(\Pi\zeta_{\rm odd})^T}_{\rm sim}$ estimated from $N_{\rm sim}$ mocks. Assuming a Gaussian distribution for $\Pi\zeta_{\rm odd}$, \eqref{eq: T2} follows a $T^2$ distribution, as discussed in \citep{2017MNRAS.464.4658S}. When combining samples, we add the $T^2$ values, with the combined theoretical distribution obtained via a convolution as in \citep{Philcox:2022hkh}, or from the empirical mock-based distribution, as above. Assuming that the fiducial and empirical covariances differ, this test will provide a different weighting than the rank-based test described above, and could thus differently highlight parity-violating signatures, though we caution that it assumes Gaussianity.

\section{Results}
\noindent We begin by discussing the empirical 4PCF covariances obtained from the \glam and \patchy suites. Whilst the correlation structures are very similar (see Fig.\,3 of \citep{Philcox:2022hkh}), the variances differ considerably, as shown in Fig.\,\ref{fig: cov-ratio}. For all data chunks except LOWZ-S, the \glam covariance significantly exceeds that of \patchy, with a roughly uniform rescaling found across the radial and angular 4PCF components. In LOWZ-S, the conclusion reverses; the \patchy variance is larger than that of \glam. These results are not unsurprising, given that: (a) \citep{Ereza:2023zmz} found a significant excess in the two-point correlation function variances of \glam compared to \patchy for the CMASS-N data-chunk, and (b) the 4PCF covariance contains contributions from the 2PCF, 3PCF, 4PCF, 5PCF and 8PCF. The above observation is central to the remainder of the paper. If the \patchy 4PCF covariance was underestimated, could this explain the excess parity-odd signal?

\begin{table}[]
    \centering
    \begin{tabular}{l||c c | c c }
     \textbf{Sample} & \multicolumn{2}{c|}{\glam} & \multicolumn{2}{c}{\patchy} \\\hline\hline
     Combined &  84.7\% & $1.4\sigma$ & 100.0\% & $>$$3.5\sigma$\\\hline
     CMASS-N & 1.4\% & $0.0\sigma$ & 98.8\% & $2.5\sigma$\\
     CMASS-S & 65.4\% & $0.9\sigma$ & 97.3\% & $2.2\sigma$\\
     LOWZ-N & 64.0\% & $0.9\sigma$ & 96.6\% & $2.1\sigma$\\
     LOWZ-S & 100.0\% & $>$$3.2\sigma$ & 95.0\% & $2.0\sigma$\\
    \end{tabular}
    \caption{Probability of a detection of parity-violation from the BOSS dataset, obtained using the rank test shown in Figs.\,\ref{fig: rank-test}\,\&\,\ref{fig: rank-test-individual}. We give the one-tailed probability of BOSS compared to the distribution of \glam (left columns, new to this work) and \patchy (right columns, matching previous work) simulations, and present also the effective Gaussian significances. Where the BOSS result is larger than all values in the empirical histogram, we give the minimum detection significance.}
    \label{tab: rank-probs}
\end{table}

In Fig.\,\ref{fig: rank-test} we present the main result of this work: a comparison of the parity-violation parameter $\tilde\chi^2$ from BOSS, \glam and \patchy. This is analogous to Fig.\,4 of \citep{Philcox:2022hkh} but now includes the LOWZ dataset, as well as the empirical distribution from the \glam catalogs. The results are striking: we find that the BOSS data has larger $\tilde\chi^2$ than every \patchy catalog but only $84.7\%$ of the \glam suite. Converted to one-tailed Gaussian significances (cf.\,Tab.\,\ref{tab: rank-probs}), this corresponds to a $>$$3.5\sigma$ excess in BOSS when analyzed with \patchy (consistent with \citep{Philcox:2022hkh,Hou:2022wfj}) or a $1.4\sigma$ deviation using \glam. If we regard \patchy as the `true' noise distribution we thus obtain a strong detection of parity-violation from BOSS, but this disappears if we assume \glam is a more realistic representation of the BOSS survey. Furthermore, we can treat each \glam mock as a `validation' dataset for the \patchy analysis: in this framework, we find $\geq 3\sigma$ detections of parity-violation in 41\% of the \glam mocks, indicating significant bias in the \patchy pipeline.

To understand these results further, we show the $\tilde\chi^2$ distributions from each BOSS chunk in turn in Fig.\,\ref{fig: rank-test-individual}, with associated numerical results given in Tab.\,\ref{tab: rank-probs}. In all cases, we find significant differences between the \glam and \patchy distributions, which occurs primarily due to the different 4PCF covariances (cf.\,Fig.\,\ref{fig: cov-ratio}). In all datasets except LOWZ-S, $\tilde\chi^2_{\rm BOSS}$ falls in the center or the left of the \glam distribution but in the tails of the \patchy distribution, matching the findings of Fig.\,\ref{fig: rank-test}. For LOWZ-S, we find the opposite results: the \glam distribution is shifted to lower $\tilde\chi^2$, thus its detection significance is \textit{larger} than that obtained with \patchy. This may indicate inaccuracies in the construction of the \patchy and/or \glam catalogs for this chunk (which has a more complex selection function than CMASS). We note, however, that this region contains the fewest galaxies, and thus the least statistical weight.

\begin{figure}
    \centering
    \includegraphics[width=0.85\linewidth]{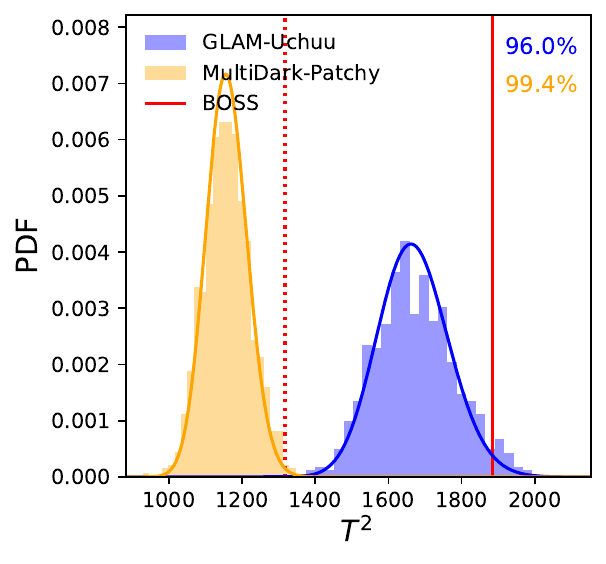}
    \caption{Alternative test for parity-violation, projecting the parity-odd statistic onto the $250$ lowest-noise eigenmodes of a fiducial covariance matrix, then computing the distribution of the $T^2$ statistic \eqref{eq: T2} using the empirical \glam or \patchy covariance. PDFs are obtained from jackknifing (histograms) and analytic Gaussian theory (lines) and we show results for all samples combined, with the numbers on the top-right giving the detection probabilities relative to the mocks. The full (dashed) red lines show the BOSS results when analyzed using the \glam (\patchy) mocks. We find no significant detection of parity-violation when using \glam but a detection when using \patchy.}
    \label{fig: proj-test}
\end{figure}

Finally, we show results from the alternative Gaussian analysis in Fig.\,\ref{fig: proj-test}. When projecting the data onto 250 eigenmodes and analyzing via the analytic $T^2$ distribution, we find a $3.0\sigma$ excess in BOSS when analyzed with \patchy or a $2.4\sigma$ detection with \glam. From the figure it is clear that the $T^2$ distribution is somewhat broader than the analytic theory, which is a signature of likelihood non-Gaussianity and leads to inflated detection significances. If one uses the empirical $T^2$ distribution rather than the theoretical, these detection significances reduce to $2.7\sigma$ (\patchy) and $2.1\sigma$ (\glam), indicating no significant evidence for parity-violation in the \glam analysis.

\section{Discussion}
\noindent What do the above results imply for the cosmic parity-violation study of \citep{Philcox:2022hkh}? The precise conclusions depend significantly on our assumptions about the \glam and \patchy simulation suites. Two choices are apparent:
\begin{enumerate}
    \item We assume that the noise distribution of BOSS is well-modeled by the new \glam catalogs but not by the older \patchy catalogs (which are less realistic, as, for example, they use an alternative method to $N$-body and do not account for redshift evolution of the galaxy samples). In this case, we find no evidence for parity-violation, with our main analysis yielding only $1.4\sigma$.
    \item We assume that neither the \glam nor \patchy suites accurately represent the BOSS noise distribution (seen, for example, by the anomalous results with the LOWZ-S data chunk). In this case, the significant variation in detection significances from two differently produced galaxy catalogs (both of which are calibrated to BOSS) indicates that our pipeline may have large systematic errors and cannot be used to claim any strong detection of parity-violation.
\end{enumerate}
In either case, we do not find significant evidence for parity-violation from the BOSS survey (using the approach of \citep{Philcox:2022hkh}), and thus no observational motivation for novel inflationary or late-time physics.

Although the statistical methods differ, this result also has implications for the analysis of \citep{Hou:2022wfj}. In this work, the main constraints come from an unprojected Gaussian analysis using the theoretical covariance matrices of \citep{Hou:2021ncj}, with the effective volume and shot-noise fitted to the empirical \patchy covariance. Although this introduces mock-based information only through a pair of parameters, one still affects significant variation in the constraints if the \glam catalogs are replaced with those of \patchy (from Fig.\,\ref{fig: cov-ratio}, given the roughly constant variance ratio). The parity-odd excess of \citep{Hou:2022wfj} is also enhanced if more bins are used in the data-vector; this would be interesting to investigate further with the \glam catalogs, though we caution that larger dimensionality is often associated with likelihood non-Gaussianity.

Finally, the analysis described in this \textit{Letter} has ramifications for future surveys such as DESI and Euclid \citep{2016arXiv161100036D,2011arXiv1110.3193L}. These experiments will lead to an explosion in the data volume, facilitating much more precise probes of the parity-violating sector (amongst many other topics). Enhanced precision does not preclude bias however, and it remains possible that model-independent analyses like the above will be limited by the accuracy of our catalogs, in particular their ability to reproduce the statistical properties of high-dimensional observables. If these hurdles can be surmounted, however, the parity-odd dataset could open doors to a wide variety of exciting inflationary studies.

\vskip 16pt
\acknowledgments
{\footnotesize
\noindent 
We thank Giovanni Cabass and Colin Hill for insightful comments. OHEP is a Junior Fellow of the Simons Society of Fellows and thanks the night markets of Taipei for their pig-related delicacies. 
JE acknowledges financial support from the Spanish MICINN funding grant PGC2018-101931-B-I00. The \glam catalogs will be made publicly available at \url{http://www.skiesanduniverses.org/Simulations/Uchuu/}.
}

\appendix

\bibliographystyle{apsrev4-1}
\bibliography{refs}

\end{document}